\documentclass[twocolumn]{IEEEtran}

\usepackage{tikz}
\usepackage{algorithm}
\usepackage[noend]{algpseudocode}
\usepackage{subcaption}
\usepackage{multirow}
\usepackage{enumitem}
\usepackage{mathtools}
\usepackage{amsmath}
\usepackage{url}

\usepackage{biblatex}
\newcommand{\fig}[1]{Fig.~\ref{fig:#1}}

\begin{document}

\title{Stop Bugging Me! Evading Modern-Day Wiretapping Using Adversarial Perturbations}
%
%
%

\author{Yael~Mathov~\IEEEauthorrefmark{1},
        Tal~Ben~Senior~\IEEEauthorrefmark{1},
        Asaf~Shabtai,
        and~Yuval~Elovici
\thanks{All authors are with the Department of Software and Information Systems Engineering, Ben-Gurion University of the Negev, Beer-Sheva 8410501, Israel}
\thanks{* The authors contributed equally.}
}

%
%

%

\maketitle

\begin{abstract} 
Mass surveillance systems for voice over IP (VoIP) conversations pose a great risk to privacy. 
These automated systems use learning models to analyze conversations, and calls that involve specific topics are routed to a human agent for further examination.
In this study, we present an adversarial-learning-based framework for privacy protection for VoIP conversations.
We present a novel method that finds a universal adversarial perturbation (UAP), which, when added to the audio stream, prevents an eavesdropper from automatically detecting the conversation's topic.
As shown in our experiments, the UAP is agnostic to the speaker or audio length, and its volume can be changed in real time, as needed.
Our real-world solution uses a Teensy microcontroller that acts as an external microphone and adds the UAP to the audio in real time.
We examine different speakers, VoIP applications (Skype, Zoom, Slack, and Google Meet), and audio lengths.
Our results in the real world suggest that our approach is a feasible solution for privacy protection.
\end{abstract}

\begin{IEEEkeywords}
    Adversarial examples, Privacy protection.
\end{IEEEkeywords}

%

\section{Introduction}
\IEEEPARstart{T}{hroughout} history, countries and their intelligence agencies have used advanced mass surveillance systems in their national security efforts \cite{cohen2010mass}.
Such mass surveillance systems collect a massive amount of data from landline communication, cellular, and endpoint devices; as well as data collected from conversations that are sent over the Internet.
During the COVID pandemic, people and companies relied heavily on the Internet for work and personal use, and communicating via voice over IP (VoIP) using third-party applications became more routine.
As a result, more people became victims of cyber attacks, which include installing spyware on the victim's endpoint \cite{interpol2020}.
Additionally, the number of nonencrypted VoIP application users (e.g., Telegram, Discord) increased dramatically due to a change in WhatsApp's privacy policy.
Therefore, an eavesdropper can collect the audio data sent over the Internet or collect it directly from a user's endpoint \cite{ibm2020what}.
The increase in the amount of data collected has led to the development of automated surveillance systems that analyze the audio samples and route conversations with sensitive content to a human agent \cite{macnish2020mass}.

At the heart of many audio-based mass surveillance systems is an automatic speech recognition (ASR) model that produces transcriptions of the recorded calls. 
The transcripts extracted by the ASR are sent to a natural language processing (NLP) algorithm to gain better understanding of the conversation.
For example, an NLP model can be used to automatically identify sensitive topics of interest (e.g., crime, terror) in a conversation \cite{lu2008opinion, hong2010empirical}. 
These models usually rely on artificial intelligence technologies, like deep neural networks (DNNs).

Despite the benefits of mass surveillance systems for society (e.g., to thwart crime and terrorism), such tools raise privacy concerns of groups and individuals.
For instance, to identify illegal activity, the data of innocent people needs to be collected and analyzed without their permission or knowledge.
Mass surveillance systems can also be utilized to achieve hostile goals. 
For example, mass surveillance can be used by business rivals to perform industrial espionage or by terror organizations to spy on civilians.
Since, many ARS systems are based on deep learning models, which have been shown to be vulnerable to adversarial perturbations \cite{szegedy2013intriguing}, this vulnerability may serve as the basis of a means of protecting users from privacy infringement.

Although some studies have explored the use of adversarial examples in the audio domain \cite{carlini2018audio, schonherr2018adversarial, khare2018adversarial}, they were not aimed at providing a solution for privacy protection in real-world settings, that post challenges to the development of effective solutions in real life.
One challenge relates to the content of a conversation, which is usually unknown in advance; as a result, such knowledge cannot be used to craft the perturbation.
Additionally, a perturbation with a constant length cannot be added to audio samples of different lengths, especially for the live data streams commonly used in VoIP applications.
Another challenge pertains to the perturbation's volume. 
The louder the perturbation is, the better it can fool the model; however, adding a loud perturbation to a conversation can prevent the listener from understanding the speaker. 
Balancing this tradeoff could be done by controlling the perturbation's amplitude, however changing the amplitude would necessitate the creation of a new perturbation; crafting one for every possible amplitude would be time-consuming and thus decease the applicability of the solution.
Additionally, in the real world, there is asymmetry between the attacker and defender; to protect him/herself, the speaker must obfuscate the context of the entire conversation, whereas the eavesdropper only needs to identify one topic in the call to compromise the communication channel.
This issue should be addressed when evaluating solutions for privacy protection in the real world.
Although adversarial perturbations offer a potential means of privacy protection, challenges such as those mentioned above must be addressed when creating a viable solution. 

In this study, we suggest a privacy preservation framework that enables participants in a VoIP call to avoid detection by a mass surveillance system based on the conversation topics.
Our approach targets the ASR component of the surveillance system; thus, manipulating the input to the NLP model and fooling the entire system.
To achieve that, our framework crafts a universal adversarial perturbation (UAP) that is agnostic to both the content of the conversation and the speaker. 
Moreover, the UAP crafted is short, and thus it can be played repeatedly, allowing the perturbation to be added to audio samples of any length, including audio streams.
We demonstrate how the perturbation's amplitude can be controlled in real time by multiplying the UAP by a constant factor, allowing the speaker to control the UAP's amplitude and eliminating the need to create multiple perturbations.
We assume that if many users routinely use our framework, an automatic mass surveillance system would be unable to distinguish conversations about topics of interest from small talk.
To address asymmetry between the speaker and eavesdropper when evaluating proposed solutions, we present an evaluation metric for assessing the UAP's ability to maintain conversations' privacy that considers the asymmetry between the speaker and the eavesdropper.

After evaluating our approach in the digital space, we present a real-world privacy preservation solution for a conversation held via VoIP applications. 
This is accomplished with a trusted external device (TED) that combines a microphone and a Teensy microcontroller that adds the UAP to the audio in real time.
Using the TED, we show how two people can have a VoIP conversation that fools the ASR model (and thus the mass surveillance system itself) of a high privileged eavesdropper.
We also show that our solution can be used for different popular VoIP applications (Skype, Zoom, Slack, and Google Meet), speakers, and audio length.
The results show the feasibility of our framework as a privacy protection solution in the real world.
By publishing our code, experimental setup, and audio records \cite{ourCode}, we hope to provide a solution for individuals and groups that wish to protect themselves in the digital age, where privacy has become a luxury.

\section{Related Work}
An adversarial example \cite{szegedy2013intriguing} is a carefully crafted perturbation to an input sample that can cause a major change in a neural network's output. 
While multiple methods for crafting adversarial perturbations have been developed~\cite{goodfellow2014explaining, carlini2017towards, chen2017zoo}, most of them have focused on perturbing a single input sample and were mainly demonstrated in the image domain.
Unlike its predecessors, a universal adversarial perturbation \cite{moosavi2017universal} is a single perturbation that fools the target model when added to multiple input samples, including both seen and unseen data.

Later, adversarial examples were presented in the audio domain \cite{carlini2016hidden,carlini2018audio}, where unintelligible perturbations were added to prerecorded audio samples to fool an ASR system.
Although the results showed promise, in both cases, the adversarial perturbations were perceptible. 
In another solution proposed for this challenge,  \cite{schonherr2018adversarial} crafted imperceptible adversarial perturbations using psychoacoustic hiding, a common MP3 technique, where a signal is compressed by eliminating tones so that when they are played simultaneously, the human ear hears all of them together as one dominant tone.
The proposed attack limited the perturbation searching space to the tones that the human ear filters out, and although the attack created imperceptible perturbations, they were tailored to a specific audio sample. 
Due to the nature of psychoacoustic hiding, using it to craft a UAP is very challenging. 
While some early studies focused on white-box attacks, others explored less permissive conditions, mostly black-box settings with oracle access to the model's output \cite{taori2019targeted, khare2018adversarial}.
However, such attacks faced other limitations, such as crafting a tailored perturbation.

Other studies presented real-world adversarial perturbations, such as attacks that craft audio-based adversarial examples that can be played over the air \cite{qin2019imperceptible, szurley2019perceptual, yuan2018commandersong}. 
Both \cite{qin2019imperceptible} and \cite{szurley2019perceptual} used the expectation over transformation framework (EOT) \cite{athalye2017synthesizing} to improve perturbation robustness for real-world transformation.
They used a simulator to simulate how each audio sample would be heard when played in rooms of different sizes. 
The simulated data samples were used to craft the adversarial perturbation to improve the attack's success when evaluated with the same transformations in the real world. 
However, none of the attacks mentioned above are feasible for privacy protection in the real world, since they were implemented with prerecorded samples.
Furthermore, the attacks were also shown to be sensitive to background noise, limiting their use in the real world. 

Neekhara et al.~\cite{neekhara2019universal} crafted a UAP that fools an ASR model. 
However, given the UAP's fixed length, it cannot be used for a conversation stream or audio records of unknown and variable length.
Additionally, the attack is created and tested using scores that are based on the character error rate (CER), a metric that is not suitable for privacy preservation evaluation. 
For example, a successful attack is defined for a perturbation that causes a CER greater than 0.5 between the original and distorted transcription, however a topic detector can identify the subject of the sample based on a fraction of the conversation.
As a result, crafting a UAP for the privacy preservation task should utilize a different metric. 

Instead of exploiting the ASR model's vulnerability to adversarial examples, some studies utilized design flaws in the microphone hardware~\cite{zhang2017dolphinattack, roy2018inaudible, song2017poster}. 
Most microphones filter out frequencies that are inaudible to humans (frequencies $>$20 kHz). 
However, due to the nonlinear nature of microphone hardware, transmitting multiple inaudible tones to a microphone results in a shadow signal that falls within the range of tones audible to humans (20 Hz to 20 kHz). 
The studies above exploited this vulnerability to transmit inaudible commands for voice controllable systems (e.g., Siri, Google Assistant, Alexa). 
This was achieved by using ultrasound speakers, whose output the microphone interprets as sound that is within the range of human auditory capabilities.   
Although this approach has shown promising results, it cannot be employed for a call between two people. When the person that sends the message uses an ultrasound speaker to mask his/her speech, the microphone will interpret both his/her voice and the inaudible waveform as above the threshold of hearing. 
As a result, the person that receives the message hears a mix of sounds which is difficult to understand.

Several studies have utilized adversarial examples as a solution for privacy protection in the image domain.
In an early example \cite{sharif2016accessorize}, the authors created glasses with adversarial patterns that allow a person to avoid detection by a facial recognition system.
Later, \cite{zhang2020adversarial} proposed an end-to-end cloud solution to preserve the privacy of user portraits.
The solution targets facial-recognition-based crawlers that browse photo sharing clouds to build a user portrait dataset.
The privacy model ``PrivacyNet"~\cite{mirjalili2020privacynet} uses adversarial perturbations to modify the input in a way that allows the user to control which soft biometric attributes he/she wishes to obfuscate. 

Privacy protection techniques were also demonstrated in the audio domain, such as an encoder-decoder architecture for ASR was used to obfuscate the user's identity \cite{srivastava2019privacy}, or the end-to-end cloud-based ASR service called Preech \cite{ahmed2020preech}, a system that protects user privacy, including textual privacy.
However, since Preech requires the user to use a particular service, it is not a solution that can be implemented across multiple platforms. 
Moreover, Preech assumes offline decoding and does not support data streams and thus is less suitable for real-world VoIP applications.
To avoid detection by mass surveillance systems when communicating via telephone, \cite{abdullah2019hear} suggested a black-box attack (with access to the oracle) that removes low-intensity components of an audio sample thus creating a new sample that fools automatic voice identification (AVI) and ASR models.
The changes to the sample are imperceptible and effective against a variety of AVI and ASR systems.
While the results were promising, the attack required at least 15 queries to the target model to create tailored changes to a specific prerecorded sample.
Since the method cannot be used in real time for unknown data, it is less suitable for real-life conversations.

Our study utilizes the knowledge accumulated in prior research to supply an adversarial learning-based, real-time solution against mass surveillance. 
In Section~\ref{subsec:design_considerations} we discuss how known challenges affected the design of our framework.

\section{Framework}
\label{chap:Framework}

In this study, we assume a scenario in which two people, one \emph{Speaker} and one \emph{Listener}, want to use their endpoints (i.e., computers) to have a private conversation via a popular VoIP application.
In addition, a malicious \emph{Eavesdropper} uses a mass surveillance system to automatically collect audio data and identify the topics discussed. 
To collect the conversation data, the Eavesdropper compromises the endpoints of the Speaker and/or Listener or by performing a man-in-the-middle attack on the communication channel to intercept the VoIP call. 
The digital waveform of the call is then sent to a deep-learning-based system, with three main components: ASR, topic detector, and alert mechanism.
The ASR component uses a DNN model to create a transcription of the digital waveform. 
Then, the transcription is sent to a DNN-based topic detector that analyzes the text and outputs a list of the topics of the conversation. 
If a specific topic is identified, an alert will be raised to inform the Eavesdropper, who will assign a human agent to listen to the call. 
If the topics are not of interest to the Eavesdropper, the call is ignored. 
Additionally, it is safe to assume that the Eavesdropper sets a confidence threshold that prevents the system from raising an alert if the topic detector identifies a topic of interest with a low confidence score.
Since the system analyzes a large amount of audio data, reporting calls that the topic detector is not confident about will result in a high false positive rate, which makes the system unreliable. 
Since the alert is usually implemented using a simple set of conditions with regard to the topic detector's output, we focus our evaluation on the two complex models (i.e., the ASR and topic detector).

\subsection{Threat Model}
We assume that the Eavesdropper's goal is to automatically identify specific topics in the conversation between the Speaker and Listener.
To do so, the Eavesdropper can collect data from the victims' endpoints and the communication channel.
We also assume that the Eavesdropper collects the audio data from the victims' endpoints using software manipulation (e.g., compromising the microphone driver) and not by targeting the hardware itself.
On the other hand, the goal of the Speaker and Listener is to maintain a private communication channel.
Hence, the victims want to talk via the VoIP application without the topic of their conversation being automatically detected by the Eavesdropper's system.
The Speaker does not use a prerecorded message or recite a known transcript, and the Listener wants to understand the Speaker but is willing to tolerate some noise on the call.
Additionally, we assume that the Speaker is aware that his/her endpoint might have been compromised by the Eavesdropper and knows which ASR model is used in the surveillance system.
The assumptions regarding the Eavesdropper were chosen to show the feasibility of our solution in the face of a high privileged adversary. 
If the solution is effective under the suggested threat model, it will also be effective under more permissive settings. 

\subsection{Framework Configuration}
\begin{figure*}
    \centering
    \includegraphics[width=0.80\linewidth]{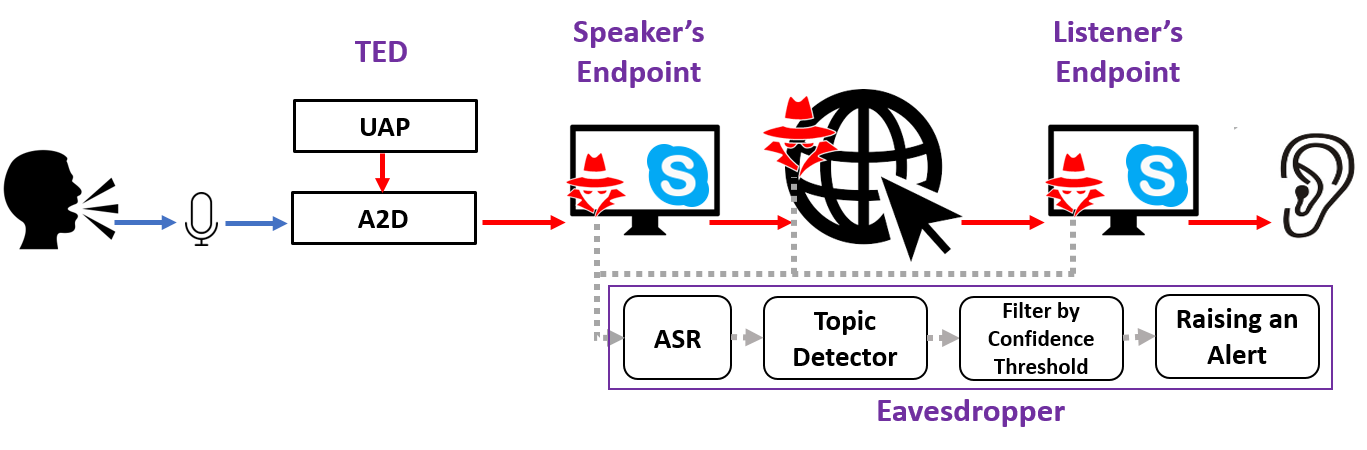}
    \caption{
    The proposed framework:
    The Speaker uses a TED as an external microphone that records his/her speech, adds a UAP to the digital audio signal, and sends it to the Speaker's endpoint.
    The perturbed data is sent, via a VoIP application, to the Listener side.
    The Eavesdropper may collect data from both endpoints and/or the communication channel.
    The collected perturbed data is sent to the Eavesdropper's surveillance system for analysis.
    }
    \label{fig:FrameworkConfiguration}
\end{figure*}

We propose the framework presented in \fig{FrameworkConfiguration} for the scenario of a private conversation between two parties described above.
The Speaker uses the knowledge he/she has on the Eavesdropper's system to craft a UAP that can fool the ASR model.
In addition, the Speaker uses a trusted external device as the input device (i.e., microphone) for his/her endpoint. 
During voice acquisition, the TED adds the UAP to the audio signal, and the perturbed audio is sent as digital audio input to the Speaker’s endpoint. 
As a result, all of the audio signals that pass through the Speaker’s computer contain the perturbation. 
The Speaker uses a third-party VoIP application to communicate with the Listener. 
The Listener can understand the content of the perturbed audio, even though there is some audible noise in the conversation. 
We assume that some or all of the audio signals are automatically collected by the Eavesdropper’s system. 
Those audio samples are fed to an ASR model, and the transcript produced is fed to a topic detector that searches for topics of interest in the conversation. 
If the topic detector's confidence in its prediction is higher than a predefined threshold, the system raises an alert that informs the Eavesdropper to manually check the conversation.
However, because the UAP was added to the digital audio signal before it was sent to the Speaker's endpoint, the ASR model produces an incorrect transcript of the audio. 
As a result, the topic detector does not receive the correct transcript and thus fails to raise an alert on the conversation's topic.

\subsection{Design Considerations}\label{subsec:design_considerations}

Our solution aims to address the following objectives:

\textbf{A real-time solution:} 
The solutions proposed in many prior studies craft adversarial perturbations for a known digital audio signal, add the perturbation to the input signal, and feed the perturbed signal to the learning model.
However, such a solution will not work for a normal, ongoing conversation in real life. 
Crafting an adversarial perturbation based on the Speaker's audio file is impossible, since in an actual conversation this data is unavailable in advance (i.e., the Speaker has not spoken, so the sample does not exist).
We address this challenge by designing a universal adversarial perturbation that can be added to any digital audio signal.

\textbf{A live data stream:} 
An adversarial perturbation for a prerecorded audio sample has a constant length. 
Hence, it cannot be used for digital audio signals of a different length or for a live audio stream. 
To address this challenge, we crafted a short perturbation that can be added repeatedly over a longer audio sample. 
As a result, the perturbation can be added to samples of different lengths, including live data streams.

\textbf{Effective against a high privileged eavesdropper:}
In 2020, both the popularity of nonencrypted VoIP applications and the number of cybercrime victims increased~\cite{interpol2020, ibm2020what}.
Therefore, a realistic solution must be effective against a high privileged eavesdropper that can collect data from the victim's endpoint or by performing a man-in-the-middle attack on the communication channel.
If the perturbation is added after the Eavesdropper collects the audio signal, it is useless. 
Therefore, the perturbation is added to the Speaker's digital audio signal in a TED. 
It is safe to assume that even if the Eavesdropper has compromised the Speaker's endpoint, he/she cannot compromise the TED connected to it. 
Thus, the digital audio signal that reaches the Speaker’s endpoint already includes the UAP. 
Such a setup decreases the Eavesdropper's ability to automatically detect the presence of the UAP.

\textbf{Agnostic to third-party applications:} 
We assume that the Speaker and Listener are regular people and thus communicate via popular VoIP applications.
Thus, our solution aims to work even when the audio signal is sent via various popular third-party applications that use common automatic compression processes (e.g., Skype, Zoom, Slack, and Google Meet).

\textbf{Blend into the crowd:} 
We assume that if many users use the proposed framework, the Eavesdropper cannot build an automated mitigation method against the vast amount of abnormal transcriptions.
Even if the Eavesdropper can develop a method to detect perturbed calls, the system will fail to automatically identify calls about the topics of interest, and the conversations have to be examined manually, which is not a feasible solution for mass surveillance.

\section{Crafting the Audio-Based UAP}
Our method aims to create one untargeted universal adversarial perturbation that can fool an ASR model.
The UAP can be added to any audio sample to cause the model to output an incorrect transcription. 
Additionally, our method ensures that the amplitude of the UAP remains smaller than a given constraint.
To support the ability to fool samples of different lengths, we define a short perturbation and concatenate it several times according to the sample's length. 

\subsection{Problem Formulation}
Let $M$ be a DNN-based ASR model that receives audio samples as input and outputs a transcription of the audio file.
For a dataset of audio samples $X$, we aim to find a UAP $\delta$ that fools $M$ when added to most samples in $X$.
Furthermore, to control the volume of the UAP, its amplitude (under $\ell_\infty$ norm) cannot exceed a given $\lambda$.
Therefore, as done in~\cite{moosavi2017universal}, for dataset $X$ and amplitude constraint $\lambda$, we search for the following $\delta$:
\begin{equation}
\begin{gathered}
    M(x+\delta) \neq M(x) \ \textbf{for most} \ x \in X \\
    \textbf{s.t.} \quad \|\delta\|_\infty\leq \lambda 
\end{gathered}\label{eq:formulation}
\end{equation}

\subsection{Repeatable Perturbation}
To create a UAP that can be used for streaming tasks, our method aims to find a UAP $\delta$ that can be added to audio samples of any length. 
We achieve this by creating a short perturbation that can be played repeatedly over longer audio samples.
However, we must ensure that the perturbation is able to fool the ASR model when added to different parts of the sample.
To do so, we integrate the replaying of the UAP as part of the perturbation's crafting process.

Therefore, we define $\delta$'s length to be $l$ seconds.
For a chosen constant $k$, we trim the samples in $X$ such that the length of each $x \in X$ equals $l \cdot k$ seconds.
To add $\delta$ to the samples, we define $\delta^k$, a $k$ time concatenation of $\delta$.
As a result, Equation~\ref{eq:formulation} is updated to find a UAP $\delta$ as follows:
\begin{equation}
\begin{gathered}
    M(x+\delta^k) \neq M(x) \ \textbf{for most} \ x \in X \\
    \textbf{s.t.} \quad \|\delta\|_\infty\leq \lambda 
\end{gathered}\label{eq:repeatable}
\end{equation}

\subsection{Objective Function}
Based on Equation~\ref{eq:repeatable}, we define an objective function $\mathcal{L}$ for finding $\delta$.
Function $\mathcal{L}$ is split into two parts: fooling the model and minimizing the UAP amplitude.
To fool the model, we use $M$'s loss function $\mathcal{L}_M$ to maximize the distance between the true output and the result for a perturbed sample.
To maintain a UAP with a small amplitude, we minimize the Euclidean norm of $\delta$.
Inspired by~\cite{carlini2017towards}, we use a constant $c$ to balance the tradeoff between the two parts.
Therefore, for a given $X$, we aim to minimize:
\begin{equation}\label{eq:loss_function}
    \mathcal{L}(X, \delta) = c \cdot \| \delta \|_2 - \mathcal{L}_{M}(M(X), \ M(X + \delta^k)) 
\end{equation}

\subsection{Optimizing the Objective Problem}
To solve the objective function in Equation~\ref{eq:loss_function}, we employ the basic iterative method~\cite{kurakin2016adversarial}.
In each iteration, we calculate the loss gradients of Equation~\ref{eq:loss_function} with respect to $\delta$.
Then, we multiply the sign of the gradients by a learning rate parameter $\alpha$ and add the result to the current $\delta$.
To improve the convergence time of the method, we start the process with a high $\alpha$ and gradually decrease it.
Finally, we clip $\delta$ to ensure that $-\lambda \leq \delta \leq \lambda$.
Therefore, finding $\delta$ can be done using:
\begin{equation}
  \begin{aligned}
    & \delta_{0} = 0  \\
    & \delta_{N+1} = Clip_{\lambda}\{\delta_{N} + \alpha \cdot sign(\nabla_\delta \mathcal{L}(X, \delta)) \} 
  \end{aligned}\label{eq:update_step}
\end{equation}

\subsection{Algorithm}
We use Equations~\ref{eq:loss_function} and~\ref{eq:update_step} to define Algorithm~\ref{alg:uap_crafting} for crafting an audio-based UAP for an ASR model $M$. 
In line 1, we start by initializing the UAP $\delta$, of size $l$, with zeros. 
As part of the initialization, we also split $X$ into batches to further improve the method's convergence time.
We chose to use random batches with audio samples from different speakers, since this improves the UAP's robustness to speaker changes.
Then, we perform $E$ epochs to craft $\delta$ (line 2), such that in each epoch we iterate over all of the batches $X_B$ in $X$ (line 3) and update $\delta$ according to Equation~\ref{eq:update_step}.
For each batch $X_B$, we start by concatenating the current $\delta$ $k$ times to create a $l \cdot k$ long perturbation $\delta^k$ that can be added to any $x \in X_B$ (line 4).
In line 5, we calculate the objective function presented in Equation~\ref{eq:loss_function}.
Line 6 updates $\delta$, as shown in Equation~\ref{eq:update_step}, but without the clipping.
After updating the UAP, we check to see whether the objective function $\mathcal{L}$ has stopped improving.
If it has, we decrease the learning rate $\alpha$ by one (lines 7-8).
At the end of each epoch (line 9), we apply clipping to $\delta$ to ensure that $-\lambda \leq \delta \leq \lambda$.
We found that clipping after each epoch achieves better results than clipping at the end of each batch.
The algorithm outputs an audio-based UAP $\delta$.
Additionally, we found that $k>2$ creates a UAP that is effective when applied repeatably on a long audio sample.

\begin{algorithm}[t]
    \begin{flushleft}
                \textbf{Input:} Training data $X$, initial learning rate $\alpha$, number of epochs $E$, UAP repetitions $k$, amplitude constraint $\lambda$, objective function constant $c$
                    
                \textbf{Output:} Universal adversarial perturbation $\delta$
    \end{flushleft}
        \caption{Crafting Audio-Based UAP}\label{alg:uap_crafting}
    \begin{algorithmic}[1]
        \State $ \delta \leftarrow 0 $
        \For{$i \leftarrow 1$ up to $E$}
            \ForAll{batch \ $X_B$ \ in \ $X$}
                \State $ \delta^k \leftarrow Tile(\delta, \ k) $
                \State $ \mathcal{L} \leftarrow - \mathcal{L}_{M}(M(X_B), \ M(X_B + \delta^k)) + c \cdot \| \delta \|_2 $
                \State $ \delta \leftarrow  \delta + \alpha \cdot sign(\nabla_\delta \mathcal{L})  $
                \If{$\mathcal{L}$ stops improving}
                    \State $ \alpha \leftarrow \max (1, \ \alpha - 1) $
                \EndIf
            \EndFor
        \State $\delta \leftarrow Clip_{\lambda}(\delta)$
        \EndFor
    \State \textbf{return} $\delta$ 
    \end{algorithmic}
\end{algorithm}

\section{Evaluation Setup}
\label{chap:universal_adversarial_perturbation}

\subsection{Models}
As described, the surveillance system has three components: an ASR, topic detector, and alert mechanism. 
While the alert mechanism can be implemented using a rule-based code, the first two components are more complex and usually use a DNN architecture.
In this study, we focus on the following models:

\subsubsection{ASR model}
We use open-source Deep Speech v0.4.1~\cite{hannun2014deep} as the ASR model. 
The input for the model is a 16-bit 16 kHz mono signal, and it outputs the transcript of the audio sample. 
To do so, Deep Speech uses Mel-frequency cepstral coefficient preprocessing and a neural network architecture with a combination of dense and LSTM layers. 
The model was trained by optimizing the connectionist temporal classification (CTC) loss, which we utilize as well. 

\subsubsection{Topic detector}
For the topic detection model we use the TextRazor API~\cite{textrazor}. 
The model receives transcripts as input and employs state-of-the-art NLP methods to analyze and extract information from the text.
When given a transcript, TextRazor outputs pairs that consist of a topic and its confidence score.
For each topic, the confidence score is a number between zero and one that reflects how sure the model is of its prediction.
The TextRazor documentation~\cite{textrazorDoc} recommends using a confidence threshold to avoid a high false positive rate. 
While TextRazor can be used for tasks like entity recognition, spelling correction, and more, we use topic tagging to perform topic detection on the output of Deep Speech.

\subsection{Datasets}
\subsubsection{LibriSpeech}
LibriSpeech \cite{panayotov2015librispeech} is a corpus of approximately 1,000 hours of English speech and transcripts derived from the LibriVox project's audiobooks. 
The data contains audio samples from different speakers, including both clean and noisy records.
Since LibriSpeech mainly contains short audio samples, it is less suitable for the topic detection task.
As a result, we only use this dataset to craft the perturbations.

\subsubsection{IBM Project Debater}
IBM's Project Debater \cite{ibmdebaterproject} is a learning-based system that can participate in debates with people on complex subjects from various domains. 
The project’s team released free datasets, including the Debate Speech Analysis dataset \cite{mirkin2017recorded} used in this study. 
The dataset contains 60 debate records from 10 different people in 16 different domains. 
We use this dataset to evaluate our framework, since each sample contains a long audio file, a transcript, and the debate category.  

\subsection{Evaluation Metrics}

\subsubsection{Word error rate}
The word error rate (WER) measures the distance between two transcripts according to the number of actions needed to turn one text into the other (i.e., insertion, deletion, or substitution).
The WER is normalized by the number of words in the first transcript, thus the error rate can be greater than one (i.e., when the number of actions is greater than the length of the reference transcript).
Since the topic detector analyzes the sentence words to detect the topic, we found that it is more important to understand the changes at the word level than the character level.
We note that evaluating a UAP for the privacy protection task should be done by examining the complete surveillance system (i.e., the topic detector).
However, we report the WER because it is a common metric for ASR models.

\subsubsection{Signal-to-noise ratio}
The signal-to-noise ratio (SNR) is a measurement used to quantify the level of a signal relative to the level of the background noise. 
Usually, as done in this work, the ratio is measured in decibels ($dB$). 
To evaluate the perturbation's loudness, we measure the UAP's distortion $P_\delta$ with respect to the original audio sample $P_{s}$ as $SNR_{dB} = 10 \cdot \log_{10}{({P_{s} \backslash P_\delta})}$.
We use the SNR to evaluate the perturbation's distortion with respect to the Speaker's signal in the real world.
A higher SNR ratio indicates a lower distortion.

\subsubsection{Privacy level}
There is asymmetry between the Speaker and the Eavesdropper: while the Speaker must conceal all of the conversations to maintain his/her privacy, the Eavesdropper needs to identify one sensitive topic to compromise the entire communication channel.
As recommended in TextRazor's documentation~\cite{textrazorDoc}, it is safe to assume that the Eavesdropper tries to avoid false positive alerts by defining a confidence threshold $\tau$ for the automatic surveillance system; thus the system will only raise an alert if the topic model's confidence in its prediction is higher than $\tau$.
Therefore, we define a perturbed conversation as private only if the Eavesdropper failed to detect \textit{all} of the conversation's topics. 

When given a confidence threshold $0 \leq \tau \leq 1$ and a clean audio sample $x \in X$, we calculate the audio privacy level of $\widetilde{x}=x+\delta$ ($APL_{\tau}(x, \delta)$) as follows: 
First, we calculate $T(x)$ and $T(\widetilde{x})$ -- the list of topics $t$ and confidence scores $cs$ that the topic detector outputs for $x$ and $\widetilde{x}$ respectively. 
Then, we filter out topics with a confidence lower than $\tau$ (the confidence threshold): $T_{\tau}(x)= \{ (t, cs) \in T(x) | cs \leq \tau \}$.
Finally, we define $APL_{\tau}(x, \delta)=1$ if at least one topic $(t, cs) \in T_{\tau}(x)$ also appears in the filtered topic list of the perturbed data $T_{\tau}(\widetilde{x})$, or $APL_{\tau}(x, \delta)=0$ otherwise. 
Hence, if the Eavesdropper can identify at least one of the original topics of the conversation, the conversation is not private.
We also define the perturbation privacy level $PPL_{\tau}(X, \delta)$ to be the percentage of $x \in X$, such that $APL_{\tau}(x, \delta) =1$ (i.e., the percentage of non-private samples in $X$).
We note that for the threshold $\tau_1 > \tau_2$, $PPL_{\tau_1}(X, \delta) \leq PPL_{\tau_2}(X, \delta)$, since there are probably more topics with confidence higher than $\tau_2$; therefore, a longer topic list increases the chance that at least one topic appears in both the clean and perturbed topic lists.

\subsubsection{Recall}
To compare the privacy level to a known metric, we also use Recall@10. 
To calculate it, we feed the clean input into the system and obtain a list of the top 10 topics. 
Then, we add the perturbation to the sample, feed it to the model, and calculate the percentage of topics that appear in both the clean and perturbed topic lists.
This metric ignores the confidence threshold and the asymmetry between the Speaker and the Eavesdropper, but it is a common evaluation metric.

\subsection{Experimental Setup} 
We use Algorithm~\ref{alg:uap_crafting} to craft a UAP $\delta$ based on a subset of 2,500 audio files randomly sampled from the LibriSpeech dataset.
Due to the short length of the samples in this dataset and our desire to concatenate the UAP more than twice, we craft $\delta$ with a length of $l=3$ seconds.
Therefore, each audio sample is cropped into a nine-second sample, such that when the UAP is concatenated three times, the resulting $\delta^3$ can be added to it.
We use Deep Speech as the ASR model $M$ and the CTC loss as the model's loss function $\mathcal{L}_{M}$ for the objective function (Equation~\ref{eq:loss_function}), and set the tradeoff parameter $c$ to $10^{-4}$.
To minimize the objective function, we split the dataset into batches of 16 samples and perform 100 epochs using the Adam optimizer.
The learning rate $\alpha$ is initialized to 10 and is gradually deceased to one when the objective function stops improving.

As done in~\cite{neekhara2019universal}, we create six UAPs that differ by their amplitude constraint $\lambda$: $\delta_{70}$, $\delta_{100}$, $\delta_{150}$, $\delta_{200}$, $\delta_{300}$, and $\delta_{400}$, with the $\lambda$ values of 70, 100, 150, 200, 300, and 400 respectively.
The six perturbations are used to examine if our approach can improve the Speaker's ability to protect his/her privacy against the Eavesdropper with respect to the UAP's amplitude constraint. 
In addition, we compare each of the UAPs to two types of random perturbations with the same amplitude constraint: \textit{random integer} and \textit{random edge}.
The random integer is uniformly sampled in the range of each $\lambda$, while the random edge is uniformly sampled with only two values: the maximum and minimum constraint.
Therefore, for $\delta_{70}$, the corresponding random integer and random edge are sampled from $[-70,70]$ and $\{ -70,70 \}$ respectively.

We start by evaluating our approach in the digital space: the perturbations are added to the digital audio samples and fed directly to the ASR model, which sends the transcript produced to the topic detector model.
Then, we implement the setup in \fig{FrameworkConfiguration} and perform a real-world evaluation.
Since the samples in LibriSpeech are too short for topic detection, we use the Project Debater dataset for evaluation.
We compare the UAP to the random perturbation using the $PPL_{\tau}$ and Recall@10 metrics.
After manually inspecting TextRazor's output on the Project Debater transcripts, we believe that a real-world mass surveillance will use a very high confidence threshold ($\tau \in [0.98, 1]$) to avoid the high false positive rate that could be caused by a large amount of audio data.
However, to allow the scientific community to validate our work, in this study, we present the $PPL_{\tau}$ for $\tau \in [0.8, 1]$.
Additional information, including the audio records, are available at~\cite{ourCode}.

\section{Evaluating the UAP in the Digital Space}
\label{chap:results}
We evaluate the UAPs in the digital space, where the perturbations are added to the raw digital audio sample which is then fed directly to Deep Speech.
The evaluation allows us to report on the UAPs' effect on each component of the surveillance system, without the potential influence of real-world noise.
We compare each of the UAPs $\delta_{70}$, $\delta_{100}$, $\delta_{150}$, $\delta_{200}$, $\delta_{300}$, and $\delta_{400}$ to the random integer and random edge perturbations.
To evaluate the perturbations' effect on the ASR model, we use the WER with respect to the perturbations' amplitude constraint.
The Speaker tries to fool the complete surveillance system (i.e., the topic detector's output, as shown in \fig{FrameworkConfiguration}), thus we use the $PPL_{0.95}$, $PPL_{0.8}$, and Recall@10 to evaluate the perturbations' influence on the system. 

\begin{figure*}[t]
    \centering
    \subfloat[WER]{\label{fig:digital-wer}
        \includegraphics[width=0.24\linewidth]{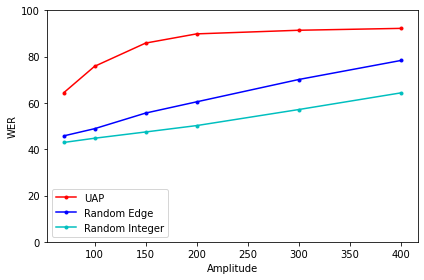}}\hfill
    \subfloat[$PPL_{0.95}$]{\label{fig:digital-ppl95}
        \includegraphics[width=0.24\linewidth]{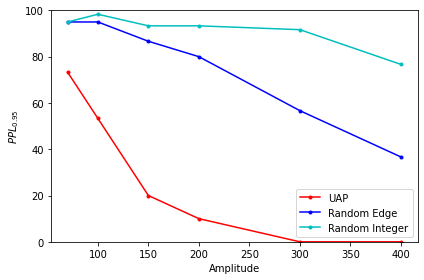}}\hfill
    \subfloat[$PPL_{0.8}$]{\label{fig:digital-ppl8}
        \includegraphics[width=0.24\linewidth]{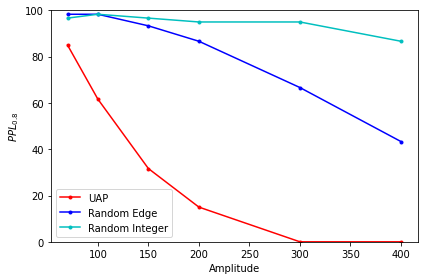}}\hfill
    \subfloat[Recall@10]{\label{fig:digital-recall}
        \includegraphics[width=0.24\linewidth]{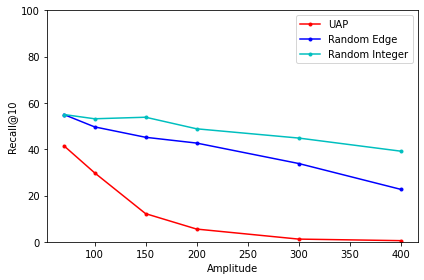}}
\caption{
    A comparison between the UAP and two random perturbations with different amplitude constraints with respect to the (a) WER, (b) $PPL_{0.95}$, (c) $PPL_{0.8}$, and (d) Recall@10 when tested on the Project Debater dataset.
    }
\label{fig:digital}
\end{figure*}

\fig{digital} presents the effects of the UAPs and random perturbations with respect to the amplitude constraints of the perturbations.
The results suggest that an increase in $\lambda$ results in an improved ability to fool the surveillance system and that the UAPs outperform the random perturbations for all amplitudes. 
Of the two random perturbations, it seems as though the random edge performs better.
This could stem from the perturbations' amplitude; random edge was designed to have values greater than (or equal to) the values of the random integer.
As expected, a greater amplitude results in greater distortion of the sample and thus has more effect on the surveillance system's output.

\fig{digital-wer} shows that the WER calculated for the UAPs are dramatically higher than the WER for the random perturbations.
The UAPs are trained using the LibriSpeech dataset but are evaluated on the Project Debater dataset; thus we can conclude that the UAPs limit the ability of the ASR model to transcribe the unknown audio samples.
Moreover, the results in \fig{digital-ppl95} and \fig{digital-ppl8}, which reflect the perturbations' effect on the topic detector, show that the UAPs fool the entire system. 
For $\delta_{150}$, the topic detector can correctly identify less than $30\%$ of the topics; for $\delta_{200}$, it drops to less than $20\%$; and for $\delta_{300}$ and $\delta_{400}$, it drops to $0\%$.
Hence, when adding $\delta_{300}$ or $\delta_{400}$ to each of the samples in the Project Debater dataset, the topic detector fails to detect even one topic that exists in both the clean and perturbed topic lists.
The Recall@10 results presented in \fig{digital-recall} support those findings.
Furthermore, the PPL results show that when the Eavesdropper uses a higher confidence threshold to avoid false alerts, preserving the conversation's privacy becomes an easier task for the Speaker.
Therefore, knowledge about the Eavesdropper's system configuration can be used to choose the best $\lambda$ value.

As shown in \fig{digital}, an increase in the amplitude of the UAP results in a decrease in the surveillance system's ability to output the correct topic list.
However, a higher amplitude causes greater distortion, resulting in more background noise, and thus limits the Listener's ability to understand the content of the conversation.
Therefore, this tradeoff should be considered when the Speaker is crafting the UAP. 
As a solution, the Speaker can control the conversation's privacy level by using multiple UAPs with different amplitude constraints.

However, crafting a UAP for each amplitude constraint is time-consuming and thus less applicable in the real world where real-time solutions are required.
Moreover, the Speaker will be restricted to using UAPs that were crafted in advance.
We address this challenge by crafting one UAP and controlling its amplitude by multiplying the perturbation by a constant. 
For example, $\delta_{100}$ could be increased to have an amplitude of $200$ by multiplying the UAP by two. 
Similarly, $\delta_{100}$ can be multiplied by $0.7$ to decrease the UAP to an amplitude of $70$. 
Therefore, controlling the UAP's amplitude can be done in real time solely by choosing the constant required. 

\begin{figure*}[t]
\centering
    \subfloat[WER]{\label{fig:amplitudes-wer}
        \includegraphics[width=0.24\linewidth]{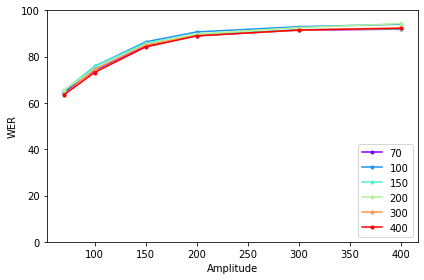}}\hfill
    \subfloat[$PPL_{0.95}$]{\label{fig:amplitudes-ppl95}
        \includegraphics[width=0.24\linewidth]{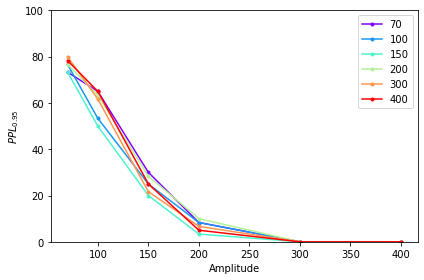}}\hfill
    \subfloat[$PPL_{0.8}$]{\label{fig:amplitudes-ppl8}
        \includegraphics[width=0.24\linewidth]{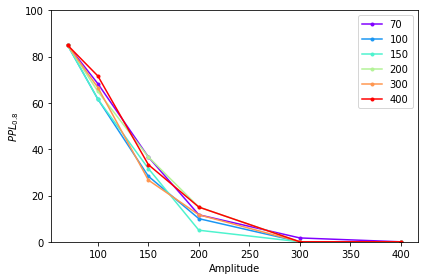}}\hfill
    \subfloat[Recall@10]{\label{fig:amplitudes-recall}
        \includegraphics[width=0.24\linewidth]{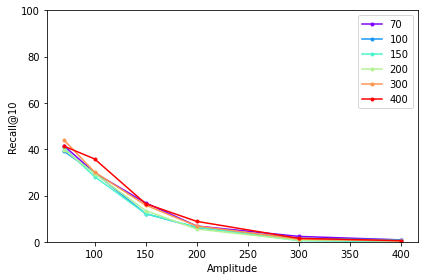}}

\caption{
    Results for the UAPs $\delta_{70}$, $\delta_{100}$, $\delta_{150}$, $\delta_{200}$, $\delta_{300}$, and $\delta_{400}$ for the Project Debater dataset.
    Each UAP is multiplied by five constants to create five perturbations with different amplitudes. 
    The graphs present the (a) WER, (b) $PPL_{0.95}$, (c) $PPL_{0.8}$, and (d) Recall@10, with respect to the perturbations' amplitude constraint.
    }
\label{fig:amplitudes}
\end{figure*}

We examine this method by multiplying each $\delta_i$ by different constants to craft UAPs for different amplitudes. 
For instance, we multiply $\delta_{100}$ by $0.7, 1.5, 2, 3,$ and $4$ to create UAPs for amplitudes of $70, 150, 200, 300,$ and $400$ respectively.
The results are presented in \fig{amplitudes}.
For each $\lambda$ value, only one UAP was crafted using this amplitude constraint, whereas the others are the result of constant multiplication.
The results are similar to those presented in \fig{digital}.
Therefore, since the UAPs maintain their ability to fool the system, we can use one UAP to create perturbations for different amplitudes.

We assume that our results reinforce the claims that adversarial examples occur in broad sub-spaces and are dependent on the direction of the gradient used to craft the perturbation and the size of the step taken in that direction~\cite{goodfellow2014explaining}. 
By multiplying the UAP by a constant, we increase the size of the step but do not change its direction, thus preserving the perturbation's ability to fool the ASR model.
A deeper analysis of this phenomenon will be performed in a future study.

\section{Privacy Preservation in the Real World}

\subsection{Real-World Implementation}
In this experiment, we implement the endpoints of the Speaker and Listener using two kits of next unit of computing (NUC) that run Ubuntu.
The parties communicate via the popular VoIP application, Skype.
To simulate the Eavesdropper's Trojan, we stream the Speaker's output audio data and the Listener's input audio data into the Deep Speech streaming model which transcribes the audio in real time.
Then, to identify the topic of the call, the transcript is sent to TextRazor.
Based on the system's output, the Eavesdropper is automatically notified of any relevant topics in the conversation.

\begin{figure}[t]
    \centering
    \includegraphics[width=0.9\linewidth]{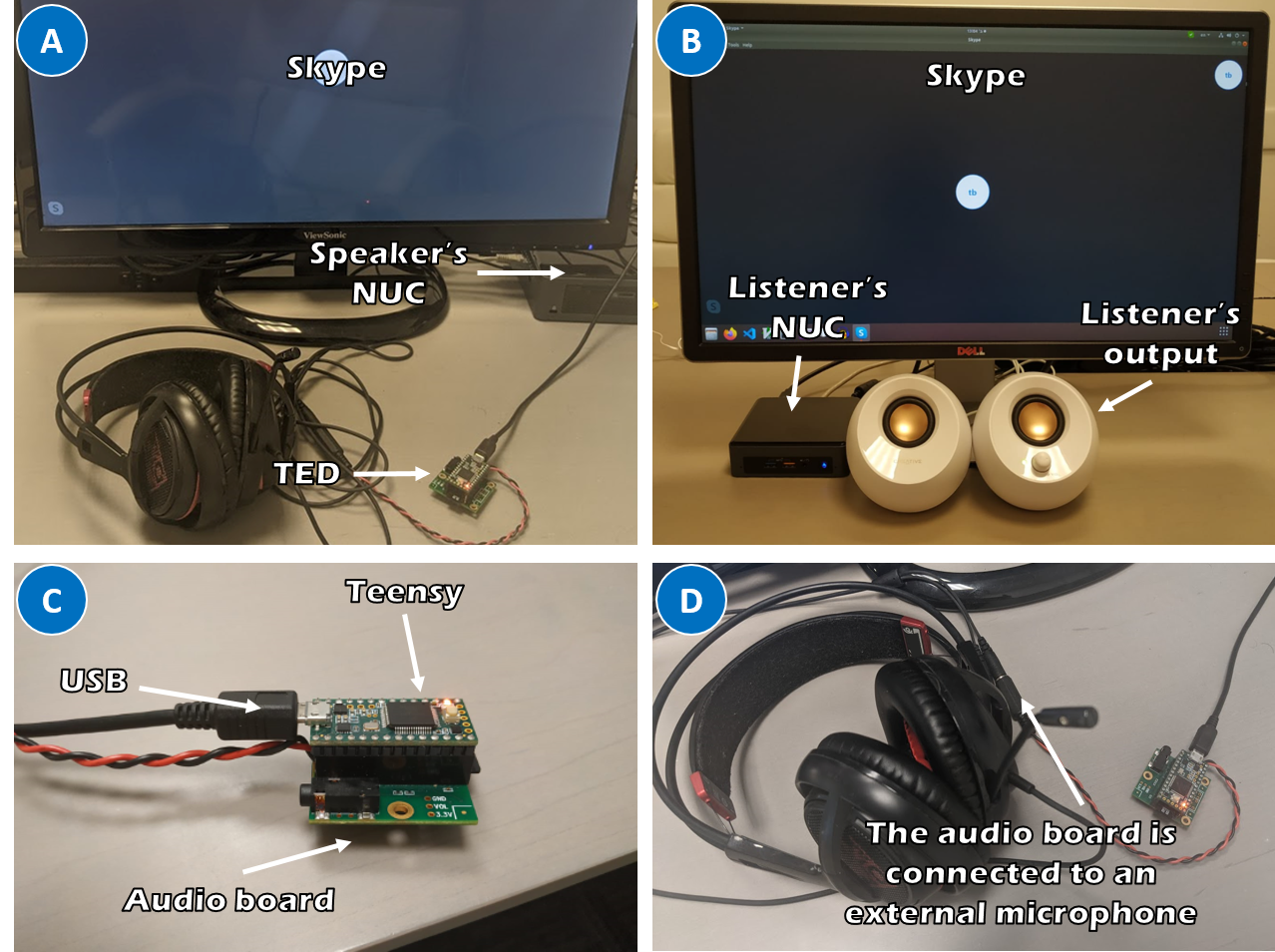}
    \caption{The implementation of the real-world setup. 
    (a) and (b) present the Speaker and Listener sides respectively. 
    (c) is a closeup of the Teensy and the audio board in the TED, while (d) shows the complete TED with the microphone.
    }
    \label{fig:app-rw-impl}
\end{figure}

To maintain a private communication channel, the Speaker uses a trusted external device as a microphone. 
The TED has three main parts (shown in \fig{app-rw-impl}): a Teensy 3.2 USB board, an audio board, and a microphone. 
The microphone is connected to the audio board via an auxiliary port, and the board is connected to the Teensy via two sets of 14 pins. 
The TED can be connected to any computer and used as an external USB microphone. 
We configured the TED as follows: The audio board samples the analog data from the microphone, converts the signal into a digital signal, and passes it to the Teensy at a 16-bit and 16 kHz sample rate. 
Then, the Teensy adds a UAP to the audio and sends it to the computer as a digital audio stream.
Finally, the perturbed sample is used for the call between the Speaker and the Listener and thus is automatically collected from their endpoints by the Eavesdropper's system.

Since we assume an eavesdropper with high privileges, maintaining the conversation's privacy relies on the use of the TED to add the perturbation.
The UAP cannot be added on the endpoint, since the Trojan can collect the clean data directly from the microphone's driver.
Also, the Eavesdropper can automatically identify the existence of the UAP by comparing the clean audio on the Speaker side to the audio collected on the Listener side.
Moreover, a privileged Trojan can also identify the process that adds the UAP and inform the Eavesdropper.
As a result, the Eavesdropper will be able to automatically filter out the perturbation from the audio.

However, by using the TED to add the UAP to the clean audio, the parties can avoid detection.
The TED is registered as a legitimate external microphone and feeds the NUC a perturbed speech signal.
Hence, the UAP already exists in all of the audio samples on both sides.
Therefore, the Eavesdropper cannot automatically detect or filter out the UAP from the clean audio sample.
As discussed earlier, our experiment assumes one Speaker and one Listener, however, supporting a conversation with two speakers (or more) can be done easily by providing an additional TED on the Listener side.

\subsection{Real-World Evaluation}
\begin{figure*}
    \centering
    \includegraphics[width=0.85\linewidth]{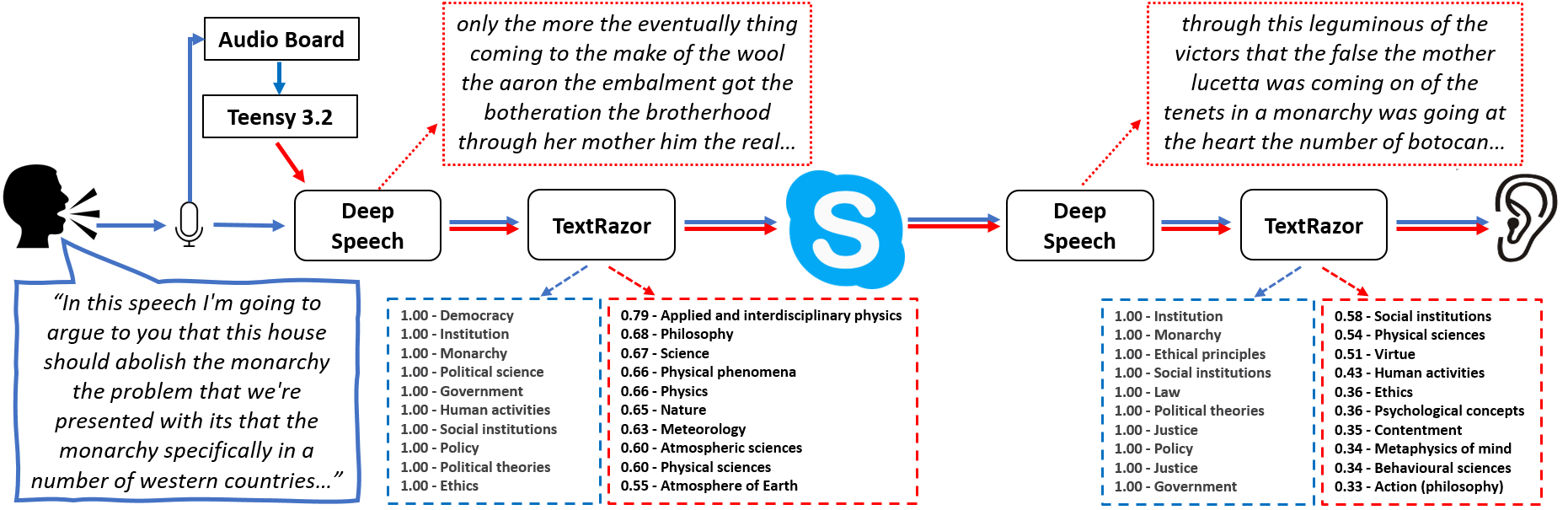}
    \caption{A real-world implementation of the framework, including an example from the Project Debater dataset.
    The blue and red arrows indicate the flow of the clean and perturbed audio respectively.
    The top boxes with the dotted red frame are the output of Deep Speech on the Speaker and Listener sides.
    The bottom boxes with the dashed blue and red frames are the output from TextRazor for the clean and perturbed transcripts respectively.}
    \label{fig:app-rw-framework}
\end{figure*}

We use the setup presented in \fig{app-rw-framework} to evaluate the UAP in the real world and compare its performance to the random edge and random integer perturbations.
For each of the three perturbations, the experiment follows similar steps:
On the Speaker side, external speakers play the audio samples from the Project Debater dataset that are recorded with the TED's microphone. 
Each sample is played twice: a clean sample without distortion and a sample in which the perturbation was added to the clean audio by the TED.
Then, the data is sent to the Speaker's NUC, which sends the audio to the Listener's NUC via Skype.
We simulate the Eavesdropper in a controlled environment in which the Deep Speech streaming model runs on the NUC on each side.
On the Speaker side, the audio is collected from the NUC's input source, while on the Listener side, the data is collected from the application's output.
On each side, the Deep Speech model receives the audio stream and transcribes it in real time.
Finally, the transcript is sent to the TextRazor API, which outputs a list of topics and their corresponding confidence scores.
Using the output for both the clean and perturbed audio samples, we measure the performance using the $PPL_{\tau}$ and Recall@10 metrics.

The results in \fig{amplitudes} show that the distortion caused by one UAP can be controlled by performing constant multiplication.
Hence, to create a UAP that has similar results to $\delta_{300}$, we can multiply $\delta_{150}$ by two in real time.
To reflect realistic use of the UAP, we choose $\delta_{150}$ as the UAP added by the TED, manually tune its volume, and report its distortion using the SNR.
A similar configuration is used for the random perturbations.
We start by evaluating our framework using the setup and configuration described above.
Then, to better understand the framework's capabilities, we perform the following set of experiments that examine the different elements in a real-world scenario: 1) change the VoIP application from Skype to Zoom, Slack, and Google Meet; 2) examine the UAP's effect on different speakers; and 3) evaluate the UAP on audio samples of different lengths.
We perform the same evaluation process using the same metrics, but we change one of the components (application, speaker, sample length) in each experiment.

\subsection{Results}
\subsubsection{Baseline}
To evaluate whether the solution is feasible in the real world, we start by performing the baseline experiment.
We send all of the samples from the Project Debater dataset to the Listener side via Skype.
We compare the UAP to the random edge and random integer perturbations and measure their distortion using the SNR.
Additionally, we evaluate the perturbations' effect on the surveillance system using the mean of the $PPL_{\tau}$ (for $\tau \in [0.8, 1]$) and Recall@10.
The metrics are computed using the data sampled on the Speaker's endpoint (NUC) before the audio has been sent via Skype and on the Listener's endpoint on the VoIP application's output.

\begin{figure}[t]
\centering
    \null\hfill
    \subfloat[Speaker]{\label{fig:baseline-a}
        \includegraphics[width=0.48\linewidth]{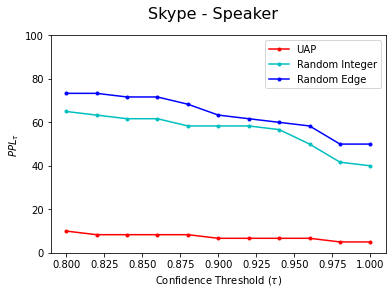}}\hfill
    \subfloat[Listener]{\label{fig:baseline-b}
        \includegraphics[width=0.48\linewidth]{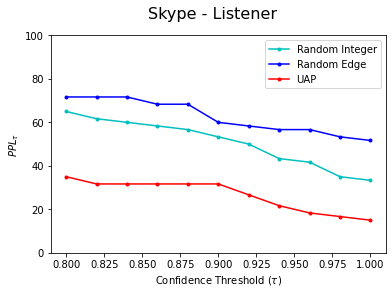}}
    \hfill\null
    
\caption{
    The PPL results for different thresholds in the baseline experiment on the (a) Speaker and (b) Listener sides. 
}
\label{fig:baseline}
\end{figure}

The results in Table~\ref{tab:rw-snr-recall} and \fig{baseline} show that the UAP is more effective than the random perturbations on both the Speaker and Listener sides.
Based on the SNR, the UAP is less than $3\%$ of the Speaker's audio but causes the surveillance system to identify at least one topic in no more than 10\% of the audio samples on the Speaker side.
However, on the Listener side, the performance of the UAP drops: for a confidence threshold $\tau \geq 0.92$, the surveillance system compromises up to 37\% of the samples; for a more realistic confidence threshold of $\tau \leq 0.95$, the system only compromises up to 20\% of the samples.
The difference between the Speaker and Listener sides might stem from Skype's data compression.
While lossless compression restores the original sample, lossy compression loses some of the data in the decompression process. 
Since Skype uses lossy compression \cite{silk}, it might delete parts of the perturbation and thus reduce the UAP's impact on the surveillance system.
The results demonstrate the importance of comparing the UAP's performance on both the Speaker's and Listener's endpoints and also show that our approach is feasible and applicable in the real world.
This experiment serves as a baseline for the experiments that follow.

\begin{table}[t]
\centering
    \begin{tabular} {|ll|c|c|c|}
        \hline
        \multicolumn{2}{|c|}{} & UAP   &  {\shortstack[c]{Random \\ Integer}}  & {\shortstack[c]{Random \\ Edge}} \\ \hline
        \multirow{2}{*}{\shortstack[l]{Skype \\ Speaker}}     & SNR       & 16.7       & 16.7      & 16.6      \\
                                           & Recall    & 10\%       & 36\%      & 38.1\%    \\\hline
        \multirow{2}{*}{\shortstack[l]{Skype \\ Listener}}    & SNR       & 17.4       & 20.6     & 20.8      \\
                                           & Recall    & 20.3\%     & 42.1\%    & 31.5\%    \\\hline
        \multirow{2}{*}{\shortstack[l]{Zoom \\ Listener}}     & SNR       & 18.7       & 22.7      & 20        \\
                                           & Recall    & 18.8\%     & 29.8\%    & 27.5\%    \\\hline
        \multirow{2}{*}{\shortstack[l]{Slack \\ Listener}}    & SNR       & 17.8       & 20.2      & 20.5      \\
                                           & Recall    & 18.1\%     & 34\%      & 32.6\%    \\\hline
        \multirow{2}{*}{\shortstack[l]{Meet \\ Listener}}     & SNR       & 19.7       & 18.5      & 18.2      \\
                                           & Recall    & 23.1\%     & 34.5\%    & 30.1\%    \\\hline
    \end{tabular}
    
\caption{The SNR and Recall@10 for all VoIP applications.
        The results obtained on the Speaker side are similar for all applications; thus we only present the results for Skype.
}
\label{tab:rw-snr-recall}
\end{table}

\subsubsection{Different VoIP applications} 
VoIP communication applications usually process the audio samples before sending them.
For instance, the application may perform echo cancellation, audio enhancement, data compression, etc.
The results for the baseline show that such preprocessing affects the perturbations' ability to fool the system.
Therefore, we examine how other VoIP applications, with different forms of preprocessing, affect our solution.
Due to their popularity during the COVID-19 pandemic for both individuals and companies, we inspect Zoom, Slack, and Google Meet to replace Skype in this experiment.
Since we only change the application, we obtain similar results on the Speaker's endpoint to those obtained for Skype; thus we focus on the results on the Listener side.

\begin{figure*}[t]
\centering
    \subfloat[Zoom]{\label{fig:rw-zoom}
        \includegraphics[width=0.3\linewidth]{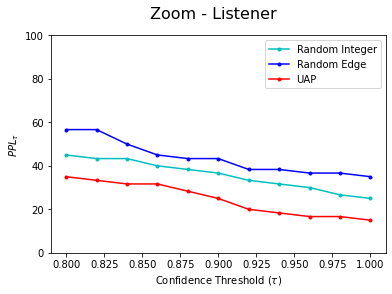}}\hfill
    \subfloat[Slack]{\label{fig:rw-slack}
        \includegraphics[width=0.3\linewidth]{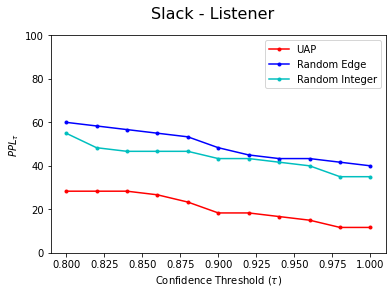}}\hfill
    \subfloat[Google Meet]{\label{fig:rw-meet}
        \includegraphics[width=0.3\linewidth]{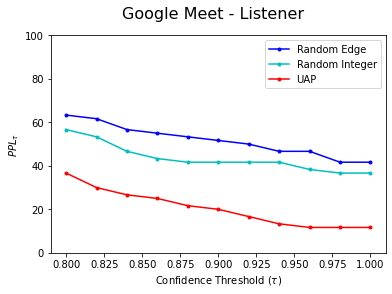}}
    
\caption{
    The PPL results for $\tau$ on the Listener side for the VoIP applications (a) Zoom, (b) Slack, and (c) Google Meet. 
}
\label{fig:rw}
\end{figure*}

The results presented in Table~\ref{tab:rw-snr-recall} and \fig{rw} support our findings in the baseline experiment with regard to the application's effect on the perturbation.
As shown by the SNR, all four applications reduce the perturbation volume on the Listener side, which results in mixed performance.
The fooling ability of the UAP drops, as shown in an increase in the $PPL_{\tau}$ and Recall@10; still, in all cases, the $PPL_{\tau}$ with a confidence threshold of $\tau \leq 0.95$ is around 20\%.
However, the random perturbations do not follow a clear pattern in the real world.
For instance, the use of Zoom causes the PPL results of the random edge to drop, but the use of Skype causes them to increase. 
While for all of the applications the random edge perturbation is more effective than random integer, in some cases the random integer performance is almost equivalent.
The results show that the effect of the applications' audio processing causes unexpected behavior by the random perturbation, thus making them less reliable for our purposes.
Additionally, for all of the VoIP applications, the percentage of compromised audio samples (PPL) is lower when using the UAP than it is when using random perturbations.
Therefore, there is an advantage in using our framework for privacy preservation in the real world.

\subsubsection{Different speakers}\label{sec:speakers} 
Past studies that used adversarial learning for privacy protection in other domains presented a tailor-made perturbation that was crafted based on the data of a specific user.
For example, since Sharif et al.~\cite{sharif2016accessorize} created the perturbation using images of a specific person, that person is the only one that can use the glasses; as a result, their method is less scalable than a user agnostic method.
In our domain, a speaker agnostic UAP means that the proposed solution will have increased accessibility.
We address this issue by creating the UAP using the LibriSpeech dataset, which contains multiple speakers of both genders.
To ensure that the UAP is speaker agnostic, we evaluate its effect on the samples of each of the 10 speakers in the Project Debater dataset.
The data contains two female and eight male speakers, each with a different speaking tone, accent, pace, style, etc.
We employ the same evaluation process and metrics used in the previous experiments, but each speaker is examined individually. 

Table~\ref{table:speaker} summarizes the results for each of the 10 speakers. 
As observed in the previous experiments, the audio on the Listener's endpoint is harder to manipulate, resulting in higher PPL and Recall@10 values.
On average, on the Speaker side, less than 6\% and 10.5\% of the samples are not private for thresholds of 0.95 and 0.8 respectively, and on the Listener side, the average $PPL_{0.95}$ and $PPL_{0.8}$ are 16\% and 32.5\% respectively.
The results highlight the difference between the $PPL_{\tau}$ and Recall@10:
For some speakers, the PPL scores are zero on both the Speaker and Listener side, but the average recall is greater than one.
These results indicate that for three speakers' audio samples (AM, DJ, and YB) there is not even one topic with confidence score over 0.8 that appears in both the clean and perturbed topic lists, yet, there are topics with lower confidence score that appear in the top places in both the clean and perturbed lists.
However, a real-world automatic mass surveillance system analyzes a large amount of data, and if it raises alerts for topics with low confidence, the system will suffer from a high false positive rate, rendering it useless to the Eavesdropper.
In contrast, a high PPL is usually reflected in a high recall; the only exception is for the speaker DZ, with $PPL_{0.8}=33\%$, which stems from the low number of audio records (in this case, one sample out of three is not private).

While on average, the results suggest that the perturbation can work for multiple speakers, there are differences in the UAP's performance for different people. 
Additionally, while the UAP preserves the privacy for speakers AM, DJ, and YB, this was not the case for JL and SN.
The other speakers showed mixed results; for example, in the case of DZ, one of the samples is not private, but the rest of them are.
We also did not find any notable differences between male and female speakers, but it is important to note that the dataset contains fewer female speakers (two females out of the 10 speakers in the dataset).
We hope to examine a larger and more diverse group of speakers in future research.

\begin{table*}[]
\begin{center}
\begin{tabular}{|c|c||c|c|c|c||c|c|c|c||c|} \hline
\multicolumn{2}{|c||}{} & \multicolumn{4}{c||}{\textbf{Speaker's Endpoint}} & \multicolumn{4}{c||}{\textbf{Listener's Endpoint}} &           \\ \hline
\textbf{Speaker ID}    & \textbf{Gender}   & \textbf{$PPL_{0.95}$} & \textbf{$PPL_{0.8}$} & \textbf{Recall@10}     &  \textbf{SNR}  & \textbf{$PPL_{0.95}$} & \textbf{$PPL_{0.8}$}  & \textbf{Recall@10}      & \textbf{SNR}   & \textbf{\#Records} \\ \hline \hline
AM  & Male     & 0        & 0       & 6.6\%     & 16.1          & 0         & 0         & 13.3\%    & 16.8      & 3 \\ \hline
DJ  & Male     & 0        & 0       & 8.7\%     & 16.7          & 0         & 0         & 7.5\%     & 17.5      & 8 \\ \hline
Dz  & Female   & 0        & 33.3\%  & 16.6\%    & 16.6          & 33.3\%    & 33.3\%    & 26.6\%    & 17.4      & 3 \\ \hline
EH  & Male     & 0        & 0       & 7.1\%     & 16.9          & 14.2\%    & 28.5\%    & 20\%      & 17.6      & 7 \\ \hline
HE  & Female   & 0        & 0       & 10\%      & 16.8          & 42.8\%    & 71.4\%    & 30\%      & 17.5      & 7 \\ \hline
JL  & Male     & 37.5\%   & 37.5\%  & 15\%      & 16.3          & 62.5\%    & 87.5\%    & 35\%      & 17        & 8 \\ \hline
SF  & Male     & 0        & 0       & 6\%       & 16.9          & 0         & 20\%      & 8\%       & 17.7      & 5 \\ \hline
SN  & Male     & 20\%     & 20\%    & 12\%      & 16.9          & 20\%      & 60\%      & 20\%      & 17.6      & 5 \\ \hline
TL  & Male     & 0        & 12.5\%  & 7.5\%     & 16.8          & 12.5\%    & 25\%      & 11.2\%    & 17.6      & 8 \\ \hline
YB  & Male     & 0        & 0       & 11.6\%    & 17            & 0         & 0         & 30\%      & 17.8      & 6 \\ \hline \hline
\multicolumn{2}{|c||}{\textbf{Mean}} 
                & 5.7\%     & 10.3\%   & 10.1\%     & 16.7          & 18.5\%    & 32.5\%    & 20.1\%    & 17.45  &     \\ \hline
\multicolumn{2}{|c||}{\textbf{STD}} 
                & 12.8\%    & 14.9\%   & 3.6\%      & 0.29          & 21.5\%    & 31\%      & 9.9\%     & 0.31   &     \\ \hline

\end{tabular}
\end{center}
\caption{
    Results for different speakers, including the speaker's gender, number of records, and $PPL_{0.95}$, $PPL_{0.8}$, and Recall@10 for both endpoints (Speaker and Listener). 
    We also present the mean and standard deviation for each metric. 
}
\label{table:speaker}
\end{table*}

\subsubsection{Different audio lengths} 
We examine how the length of the audio sample affects the ability of the Speaker and Listener to hold a private conversation.
Does a longer conversation increase the Speaker's risk of detection, and if so, to what extent?
Therefore, we sort the records in the Project Debater dataset by their length, where the shortest and longest samples are 168 and 458 seconds respectively. 
We split the samples into seven groups based on the length of the sample, and for each group, we perform the same evaluation as in the baseline.

\begin{figure}[t]
\centering
    \subfloat[Speaker]{\label{fig:speech_len-a}
        \includegraphics[width=0.7\linewidth]{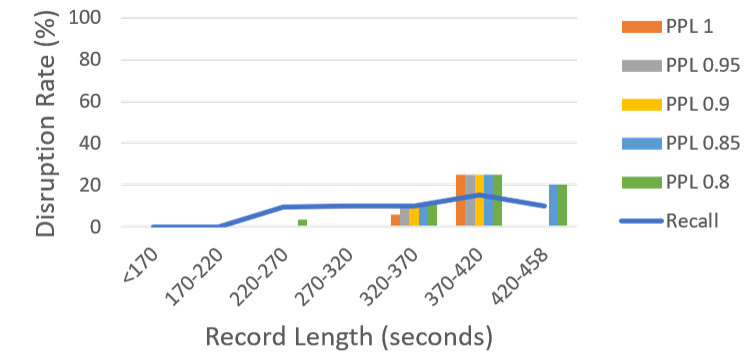}}\hfill
    \subfloat[Listener]{\label{fig:speech_len-b}
        \includegraphics[width=0.7\linewidth]{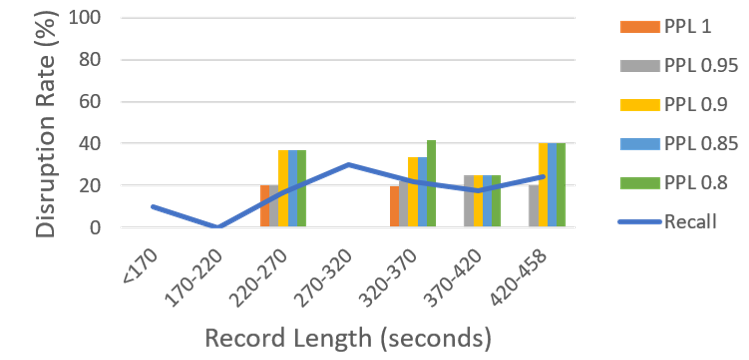}}
    
\caption{The PPL and Recall@10 results for the (a) Speaker and (b) Listener sides, with respect to audio length.
}
\label{fig:speech_len}
\end{figure}

The results presented in \fig{speech_len} suggest that, on average, the longer the conversation, the lower the percentage of private samples.
Since a longer conversation has more words that the UAP needs to manipulate to hide the conversation topic, the perturbation becomes less effective.
An exception is a middle group with audio samples that are 270-320 seconds long, which obtains a 0\% PPL for all thresholds but a relatively high recall on the Listener side; hence, the surveillance system identifies the topic but with a very low confidence score.
These results might stem from the speaker's identity and not the audio length, since this group mainly contains samples from the speaker YB.
Moreover, it seems as though the UAP ensures the privacy of both the Speaker and Listener for audio samples under 220 seconds; however, we plan to extend this experiment in the future.

\section{Discussion}

Many studies aimed at creating an imperceptible perturbation, but this criterion is not relevant in this case. 
In automated surveillance systems, it is less likely for a human to analyze the massive amount of audio data; therefore, as long as the Listener can understand the Speaker, the UAP is not required to be imperceptible.
The Eavesdropper can still utilize anomaly detection methods and language models to detect the parties trying to manipulate the conversation.
However, if many people were to consistently use the proposed framework, any attempt to use the automatic mechanism to extract the topics of interest would become futile.
For instance, if a company integrated our solution on the endpoints of all of its employees, it could ensure that all transcriptions of the company's calls are abnormal  thus protecting itself from industrial espionage.
However, using the same UAP for all endpoints increases the chances of the Eavesdropper getting his/her hands on the perturbation.
Therefore, we suggest changing the parameters of Algorithm~\ref{alg:uap_crafting} to create a different UAP for each endpoint; thus, even if the Eavesdropper identifies one perturbation, the entire company would not be compromised.

While a few studies presented defenses against input manipulation in the audio domain \cite{wang2019secure, wang2019voicepop, blue2018hello, wang2019defeating}, they are largely ineffective against our UAP. 
Defenses that detect whether the audio sample sent over the air is a human speaking or is being played through a speaker are ineffective, since our UAP is digitally added to legitimate speech. 
Furthermore, many existing defenses focus on AVI systems, while we target the less explored speech-to-text task. 
Additionally, generic defenses against adversarial perturbations have been shown to be ineffective \cite{athalye2018obfuscated}. 
Therefore, we propose a possible simple mitigation method that allows the Eavesdropper to reduce the perturbation's ability to fool the ASR model.
The UAP is a pressure wave, and thus it can be cancelled by emitting a sound wave with the same amplitude but with an inverse phase (noise cancellation). 
Since the UAP is unknown to the Eavesdropper, he/she cannot distinguish the perturbation from the background noise. 
However, we found that performing noise cancellation based on a small sample of the background audio without speech is enough to reduce most of the UAP's effects. 
Therefore, with a small reduction in the recording quality, the Eavesdropper can use noise cancellation based on the beginning or end of the audio sample to filter out most of the UAP.
To overcome this mitigation, the Speaker can use voice activity detection to play the UAP only when speaking.
Without any speechless background noise, the Eavesdropper's ability to filter out the perturbation is limited. 

\section{Conclusions}
In this study, we suggest a framework that utilizes adversarial perturbations for privacy protection in the audio domain.
We present a speaker agnostic UAP that can be added to unknown audio samples of any length and demonstrate how to control its amplitude in real time using constant multiplication.
We create a real-world implementation that uses a TED to add the perturbation to the audio stream.
Using this setup, we evaluate our solution with four popular VoIP applications, multiple speakers, and different audio lengths.
We also examine a potential mitigation that can be used by the Eavesdropper.
The COVID-19 pandemic has caused more people to communicate over the Internet, placing their VoIP conversations under the scrutiny of mass surveillance systems and increasing the need for solutions aimed at protecting their privacy. 
We believe that adversarial learning might be the key to an effective privacy preservation solution.


\appendices


\ifCLASSOPTIONcaptionsoff
  \newpage
\fi



%
\printbibliography[heading=bibintoc]

%

\begin{IEEEbiography}[{\includegraphics[width=1in,height=1.25in,clip,keepaspectratio]{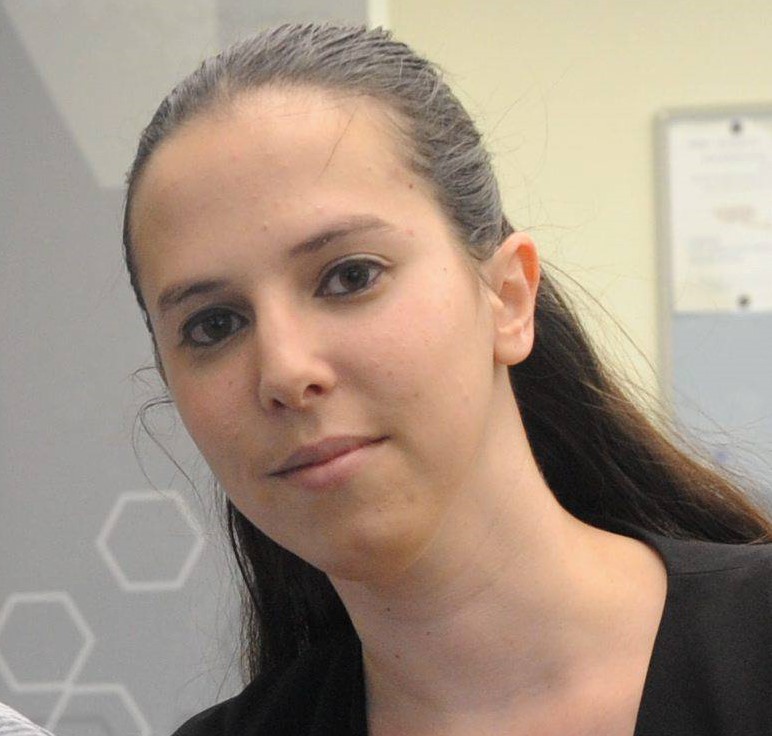}}]{Yael Mathov}
    is a data science researcher at Ben-Gurion University of the Negev (BGU) Cyber Security Research Center and a Ph.D. student of the Software and Information Systems Engineering department at BGU.
    She holds a B.Sc. degree in Computer Science and M.Sc. degree in Information Systems Engineering with a specialization in Cyber Space Security from BGU.
    Her primary research interests lie in the development and implementation of real world adversarial learning techniques.
\end{IEEEbiography}

\begin{IEEEbiography}[{\includegraphics[width=1in,height=1.25in,clip,keepaspectratio]{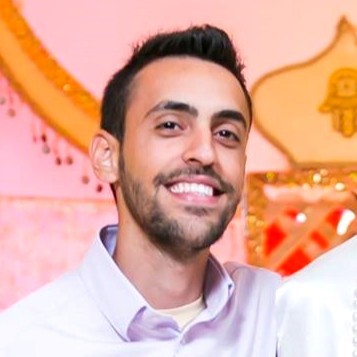}}]{Tal Ben Senior}
    is an Algorithm Engineer at Taboola. 
    He holds B.Sc. and M.Sc. degrees in Information Systems Engineering from BGU. 
    His primary research interests is adversarial learning in the audio domain.
\end{IEEEbiography}

\begin{IEEEbiography}[{\includegraphics[width=1in,height=1.25in,clip,keepaspectratio]{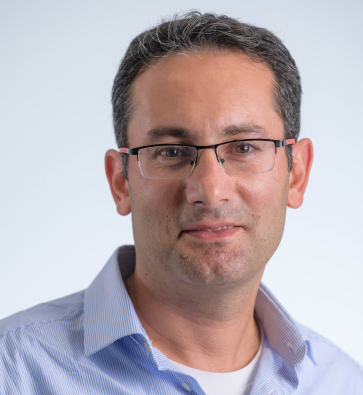}}]{Asaf Shabtai}
    is a professor in the Department of Software and Information Systems Engineering at BGU. 
    His main areas of interest are computer and network security, machine learning, and adversarial machine learning.
\end{IEEEbiography}

\begin{IEEEbiography}[{\includegraphics[width=1in,height=1.25in,clip,keepaspectratio]{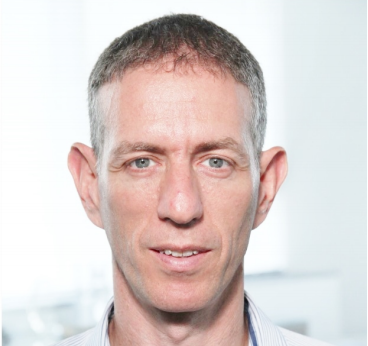}}]{Yuval Elovici}
    is the director of the Telekom Innovation Laboratories at BGU, head of BGU Cyber Security Research Center, Professor in the Department of Software and Information Systems Engineering at BGU. 
    He holds B.Sc. and M.Sc. degrees in Computer and Electrical Engineering from BGU and a Ph.D. in Information Systems from Tel-Aviv University. 
    His primary research interests are computer and network security, cyber security, web intelligence, information warfare, social network analysis, and machine learning. 
\end{IEEEbiography}




\end{document}